# A Turing-based bimodal population code can specify Cephalopod chromatic skin displays

Khalil Iskarous*
University of Southern California
kiskarou@usc.edu

Jennifer Mather
University of Lethbridge
jennifermather05@gmail.com

Jean Alupay
Greater Farallones Association
jalupay@farallones.org

*Corresponding Author:
Khalil Iskarous
1538 S. Westgate Ave.,
Los Angeles, CA 90025
kiskarou@usc.edu

## Abstract

The skin of a cephalopod forms a dazzling array of patterns made by chromatophores, elastic sacs of pigment that can be expanded by muscles to reveal their color. Thousands of these chromatophores in different parts of the skin can work together to generate a stable display of stripes, spots, mottled grainy camouflage, or dynamic oscillations and traveling waves of activation**.** How does a neuromuscular system organize the coactivation of thousands of degrees of freedom through simple central commands? We provide a minimally-complex physiologically-plausible mathematical model, using Turing's morphogenetic equations, that can generate the array of static and dynamic types of skin displays seen in several cephalopod species. The dynamical equations specify how muscle cells on the skin need to locally interact for the global chromatic patterns to be formed, and we provide a biochemical interpretation of the equations in terms of calcium dynamics. Even though the answers we provide are for the specific case of cephalopod skins, we believe that they are relevant for how global patterns arise in general brain systems of many animals that exhibit static and dynamic patterns similar to that on cephalopod skin. We also demonstrate a link between Turing neural computations and the asynchronous type of computing that has been extensively demonstrated in brain systems: population coding, using bimodal codes, with the relative heights of the modes specifying the kind of global pattern generated. Since cephalopod skins are a "visible neural net", we believe that the computational principles uncovered through their study may have wider implications for the functioning of other neural systems.

**Introduction**: One of the most spectacular aspects of the biology of the cephalopods is their variable chromatic skin displays (Hanlon & Messenger, 2018). Many animals have an external appearance that is sometimes complex and is adapted to their environment (Bard, 1981), but cephalopods are special in that they have not one but a variety of appearances (Borrelli et al., 2006). Within hundreds of milliseconds, they can produce static spots, lines, and patches of different levels of regularity (Figures 3 and 4, Column 1), allowing them to sometimes blend into their environment and at other times to increase their conspicuousness (Cott, 1940; Zylinski et al., 2011). At other moments they can exhibit dynamic traveling waves, flashing and flickering (How et al, 2017; Rosen and Gilly; 2015; Laan et al., 2014). The chromatic displays are produced by activation of radial muscles that can expand chromatophores, elastic sacs of protein of different color (Messenger, 2001). Activity of tens to hundreds of thousands of these chromatophores is necessary for accomplishment of the displays, controlled centrally from the optic lobe through the chromatophore lobe (Liu and Chiao, 2017). What computations underlie this fast pattern formation process, which requires control of many thousands of degrees of freedom? Consider, for instance, the pattern in Figure 4b on octopus, *Abdopus sp.,* which we filmed in tidal pools in Okinawa, Japan, where macroscopic foreground ovals of various sizes, organize the skin, through expansion of chromatophores with dark pigment creating a background for the ovals. This display requires coordinative computation that leads to correlation between chromatophores separated by thousands of others, to create macroscopically structured patterns. At another moment, a wave of activation, called passing cloud (Mather and Mather, 2004), could traverse the skin of the same octopus (See Video 1 in Supplementary Information), requiring computation coordinating activity in highly distant chromatophores on the edge of the wave. The possibility that central commands from the optic lobe separately instruct each individual chromatophore, or small populations of them, as to how it should contribute to the display, has always seemed unlikely due to the speed of pattern formation, the large number of chromatophores, and the enormous variation in pattern that an animal can produce. For these reasons, a different local cooperative mode of computation has been suspected at least since Hoffman (1907). Packard and Sanders (1969) and Packard (2006) argued for what became called a Horizontal Control Hypothesis (Rosen, 2016): local communication amongst muscle cells in the skin and/or neurons in the higher brain centers results in the global patterns. Packard and Sanders (1969) specifically mention a lateral inhibition local communication as the origin of pattern. And Boettiger et al. (2009) (Appendix E) and Boettiger and Oster (2009) mention the possibility that reaction-diffusion equations, or their Wilson-Cowan equivalent (Wilson and Cowan, 1972, 1973; Amari, 1977), with Turing and Hopf bifurcations, could be the kind of dynamical systems computation used to generate cephalopod displays. Turing (1952)'s pattern formation mechanism, which will be explained at length later, is relevant for the Horizontal Control Hypothesis because in such processes, the same local interactions in space and time throughout a medium, like a skin, could lead to a heterogeneous global pattern. This is despite the fact that each of the interactions, by themselves, leads to the same equilibrium value at every location in the medium, leading to the prediction of a homogeneous pattern. Turing proposed that this surprising genesis of heterogeneous patterned form from an aggregate of interactions, which by themselves would lead to lack of form, to be the route to form, or *morphogenesis*, in many areas of biology. Since then, this patterning behavior has been proven to take place at many scales of biology (Meinhardt, 2012): development (Newman et al., 2018), ecology (Meron, 2015; Malchow et al., 2008), oncogenesis (Chaplain and Sherratt, 2001), and animal skill patterns (Murray, 2001). In this paper, we demonstrate that the Horizontal Control Hypothesis can indeed by realized through the Turing mechanism. We also show that this process id physiologically plausible, and that, surprisingly, the number of control degrees of freedom to generate a large variety of static and dynamic patterns is surprisingly just *three*. Moreover, we show that these two parameters can be linked to parameters of a bimodal population code, allowing for asynchronous stochastic realization of the Turing process.

The obstacle to the development a framework to test the Horizontal Control Hypothesis is that for decades, researchers failed to find evidence for local communication between the muscles of different chromatophores (Reed, 1995). And without such local interaction, reaction-*diffusion* mechanisms are not possible, since diffusion involves spatial interaction. However, recent neurophysiological evidence (Rosen and Gilley, 2015; Rosen, 2016; Rosen et al., 2017) shows that inter-chromatophore interaction is indeed possible. And 3D electron microscopy studies of chromatophore muscles (Senft et al., 2021) show striking evidence that muscle cells of nearby chromatophores can indeed communicate through myo-myo interaction, and that even neurons rising from the chromatophore lobe have axo-axonic synapses in the skin, allowing for further conditioning of within-skin communication (Senft et al., 2021). In this paper, we build on this recent evidence to re-examine the possibility of reaction-diffusion mechanisms for online skin displays. Our first goal is to show that reaction-diffusion equations, long *suspected* to play a computational role, *can* indeed generate the repertoire of both static and dynamic patterns observed on cephalopods since antiquity, and described in modern ethograms (Hanlon & Messenger, 2018; Mather and Alupay, 2016; How et al., 2017). However, these differential equations can have many parameters. What is the smallest number of individually tunable parameters *necessary* for generating the observed repertoire? This question about the computation is important, since its answer can provide a lower bound for the number of parameters that control centers need to provide for different patterns to be realized on the skin, and could therefore provide information for experimental work probing pattern formation in the animal. The number of degrees of freedom for the motor control of this system has been argued to be low (Reiter et al., 2021), but how low? Our second goal, accomplished through numerical simulation and analytic study of the equations, is to show that necessary number of controllable parameters is three. That is, the hundreds of thousands of degrees of freedom in the skin can be controlled using a local-to-global reaction diffusion mechanism, applying uniformly across the skin (or patches thereof for pattern localized to mantle or arms for instance), with extremely few degrees of freedom, specifically three. The third goal is to propose biological implementation mechanisms that relate reaction-diffusion mathematics to what could actually be happening, biochemically and stochastically at the population level, to realize these patterns. Specifically, we propose that the rise of calcium concentration through the dynamics of Calcium Induced Calcium Release (Berridge, 2007; Bers, 2002) could be the biochemical instantiation of the dynamics of pattern formation. We do so by providing a biochemical analysis of the terms of a Substrate-depletion reaction-diffusion differential equation (Meinhardt, 1982) that can generate the repertoire. Moreover, it is known that actual computation in biological systems is highly stochastic. We therefore link Turing pattern formation to population coding of firing through simulations of associated Fokker-Planck equations (Marquez-Lago and Padilla, 2014).

Besides arguing for a computational mechanism for cephalopod skin displays, we have a broader goal, since many neural and muscular systems exhibit global patterns, which could arise, at least partially, through local interaction. Evidence has been mounting, for instance, from studies on a variety of animals, that correlations in activity can exist across space and time in brain systems, and that such extended patterns of activity play essential functional roles. For instance, Bhattacharaya et al. (2022) show that there are multidirectional traveling waves in monkey prefrontal cortex, and that changes in these spatially and temporally varying patterns correlate with working memory function (see also Zhang et al., 2018; Dahmen et al., 2022; Fernández-Ruiz et al., 2021). Also, many have shown long distance communication between large areas of the brain for binding object properties in perception, as well as linking perceptual and motor areas (Singer, 1999; Melloni et al., 2007; Assaneo and Poeppel, 2018). Further, Ermentrout and Cowan (1979), Bressloff et al. (2002), and Pearson et al. (2016) have shown that static geometric patterns can develop across the visual cortex in humans. We believe that the cephalopod body patterning system is a valuable model to use for investigating the central question of how complex spatiotemporal global patterns arise in systems, as cephalopod skins are a "a 'neural' net that can be seen with the naked eye" (Packard, 2001).

**Behavioral Patterns**: Cephalopods are capable of enormous variation in the patterns they can have on their skin, as seen in the first columns of Figure 3 and 4. Experts in the chromatic behavior of cephalopods have used terms like lines, spots, reticulate, blotch, mottle, and uniform to describe general classes of static patterns (Hanlon and Messenger, 1988; Mather and Alupay, 2016). One major axis of variation in patterns was identified by Hanlon et al. (2009): disruptiveness. Patterns high in that dimension increase the conspicuousness of the animal in its environment through the use of distinct geometric shapes like lines and spots. Lines can be relatively straight lateral bars or antero-posterior stripes (Figure 3a), or the stripes can intersect and be more curved (Figure 3b). The lines can also occur next to spots (Figure 3c). Another simple disruptive pattern is regular spots that are either light (inactivated chromatophore muscles) on dark (activated chromatophore muscles) (Figure 3d) or dark on light (Figure 3e). As irregularity increases and disruptiveness decreases, the lines can be quite short and at angles to each other, which are usually called *reticulate* (Figure 4a). The spots can also be highly irregular polygons or ovals, usually called *blotch*, and these ovals/polygons can be dark on light (Figure 4b) or light on dark (Figure 4c). As irregularities increase, the scale becomes even smaller, and we get patchy textures of variable grain (Figures 4d, 4e), usually called *mottle*. It is also possible to have the entire animal be uniform light (Figure 4f) or dark (Figure 4g). For dynamic patterns, How et al. (2017) identify several patterns which could be classified as oscillatory or traveling wave patterns. For both static and dynamic patterns, there can be an enormous amount of variation (Borrelli et al., 2006), so the patterns discussed above are really overlapping categories. A question that arises whenever one investigates a repertoire of animal behaviors with high variability is whether there are simple rules behind the patterns. We think of these patterns much like language, highly productive, in the sense that novel sentences can be generated all the time, due to the existence of a few basic principles. So the question from this linguistic point of view is whether there is a small generative device that generates the highly variable skin displays. This is the question that we ask in this paper, which we have asked before of the human behavior of speech (Iskarous, 2019), and our answer in both cases is that reaction-diffusion equations are such a finite generative device.

**Biochemical Dynamics**: As Berridge et al. (1998) have put it: "Almost everything we do is controlled by $Ca^{2+}$-- how we move, how our hearts beat and how our brains process information and store memories." And they certainly do not limit the fundamental importance of calcium signaling, to "we" as people, but to most of animal life. Indeed Packard (2006) suggested that horizontal control could be realized specifically through Calcium-based waves of activation. Calcium signaling is a possible biochemical signaling mechanism for horizontal control, since it is crucial to both neural and myogenic systems, and since cephalopod skin pattern formation could take place either in the optic and chromatic lobes and/or the skin itself, as a syncytium. Indeed, the evidence uncovered by Senft et al. (2021) of both myo-myo junctions and axo-axonic synapses suggests that the patterning could result in computations in both systems, with the common vocabulary being calcium dynamics.

The basic setup of $Ca^{2+}$ cellular biochemistry is that while the cytosol normally has extremely low concentration of $Ca^{2+}$, there is a very large concentration outside the cell, as well as inside special storage areas like the endoplasmic reticulum (sarcoplasmic reticulum in muscle cells) and mitochondria, as well as buffered by special proteins (Berridge et al., 2000). A ligand's presence outside the cell, and a sequence of events due to it, or communication from other cells through gap junctions, can introduce into the cytosol inositol-1,4,5-triphosphate receptors ($IP_3$) and/or ryanodine receptors (RyR), which can then become pores through the $Ca^{2+}$ stores releasing large amounts of it into the cytosol, which can initiate several types of events in cells or diffuse through gap junctions to other cells, spreading waves of activation (Leybaert and Sanderson, 2012). Crucially, $IP_3$ exhibits an allosteric enzymatic mechanism, where it needs calcium to be activated, so $Ca^{2+}$ exhibits an autocatalytic rise with the help of $IP_3$, but the latter needs $Ca^{2+}$ for activation. This is usually called Calcium-Induced Calcium Release (CICR), and as described by Bers (2002), is involved, in muscle cell interactions. When there is a large concentration of $Ca^{2+}$, the allosteric nature of the enzymatic mechanism, though, leads to a drop in $IP_3$ activation, once all of the sites for $Ca^{2+}$ have been occupied, leading to lesser $IP_3$ activity and lesser $Ca^{2+}$ output. One very important fact about the diffusion of $Ca^{2+}$ and $IP_3$'s diffusion is that $IP_3$ diffuses faster than $Ca^{2+}$ (Allbritton, et al., 1992).

The fundamental reaction between $Ca^{2+}$ and its facilitator/inhibitor $IP_3$, and their diffusion, is a pattern generation mechanism which operates in nature from ultraslow time scales, such as in calcium waves in the egg, to ultrafast, such as myocardium contraction, operating a billion times faster, and it can generate static structured patterns, oscillation, and traveling waves (Jaffe, 2008). Rosen and Gilley (2017) doubt the involvement of $Ca^{2+}$ waves in skin displays, since many cases of such waves are quite slow, on the order of a micrometer/second, whereas the waves required for displays would be on the order of centimeters per second. We believe that genetic evidence of a large number of genes coding for protocadherin proteins in the octopus (Albertin et al., 2015), which promote intercellular binding and adhesion, could support the idea that the skin of the octopus can act like a syncytium, easily passing activity from one cell to another (Senft et al., 2021), raising the probability that the speed of $Ca^{2+}$ waves could be high enough in this case. In addition, $Ca^{2+}$ pattern formation is involved in both muscle cell contraction, $Ca^{2+}$ the contractile machinery, as well as in neural cells, where $Ca^{2+}$ spikes are the signal necessary for neurotransmitter vesicle docking and exocytosis.

There are several very successful reaction-diffusion models of the mechanisms just described (Dupont et al., 2016; Sneyd et al., 2017; Keener and Sneyd, 2009; Falcke, 2004). They use a toolbox approach, where for each biochemical interactant, they add a differential equation, and interaction terms in other equations, and solve a large system of differential equations. This type of work is based, of course, on the legacy of Hodgkin and Huxley (1952)'s work on the squid giant axon. But there is also another tradition in computational neuroscience, where the simplest kind of equation reproducing known data is posed and analyzed for deep insight into the minimal mechanisms required to understand the phenomena. An example of this is the Fitzhugh-Nagumo proposal of 2 differential equations, instead of 4 Hodgkin-Huxley equations, to model many behaviors of neurons (Izhikevich, 2007). The latter is the approach we will take in this paper.

**Substrate-Depletion Turing Mechanism**: Turing (1952) proposed a mechanism for biochemical morphogenesis, the establishment of structural form in an initially mostly homogenous medium, with some scattered noise. His work was entirely mathematical, but in the last 2 decades, it has met with a lot of success in the laboratory (Meinhardt, 2012). In his model, the domain, or field, has two variables at each point. One variable, the *activator* (A), promotes its own growth, and the growth of the other variable, the *inhibitor* (I), which in turn inhibits the growth of A, and also inhibits itself. So there are two fields A and I, which can interact amongst each other, constituting the *reaction*. In addition, A and I can both *diffuse*. In this process a substance can spread throughout a medium, molecularly through Brownian motion. Diffusion is mathematically equivalent to a simple process where the value at each point in the field tends to become closer to the value in its local spatial neighborhood (Maxwell, 1871). Therefore, diffusion is correlating, as it can get nearby locations to have similar values of A or I. Note the *locality* of the computation in this model: the reaction of A and I occurs at each point, and the diffusion relates each point only to its neighbors. These reaction-diffusion systems can have very different functional forms (Murray, 2001). For instance, in some systems A's self-promotion or promotion of I is linear, increasing A or I equally for equal increases of A, and in other systems the reaction could be sigmoidal, with great I increases for medium magnitude increases of A, and hardly any change in I for large-magnitude increases of A. Also, given the functional forms of the reaction, the degree to which I inhibits A, or A promotes itself, for instance, is a variable reaction parameter, a coefficient. And the speed of diffusion is also a coefficient, which could be different for A and I. If a diffusion coefficient is large, the correlational tendency of diffusion extends further than if it is low.

Turing's surprising insight was to establish the mathematical conditions under which the same purely local interactions throughout the medium can result in stable structured global patterns where there is a relationship between the activator values at arbitrary spatial separation within the medium (Turing, 1952). This is surprising not only because reaction and diffusion are local, but also due to the tendency of diffusion to eliminate structure and pattern by erasing local differences. The most important of these conditions is that the speed of diffusion of the Inhibitor I is greater than that of the Activator A, i.e., that the diffusion coefficient of I is higher than that of A (other conditions will be discussed later). Under these conditions, homogeneity, which is preferred by diffusion becomes unstable, and inhomogeneity, i.e., pattern, becomes stable (Meinhardt, 1982; Murray, 2001). The physical intuition behind this crucial condition is that as A diffuses slowly, reaching some small stretch of the field, I diffusing faster and can reach beyond the span of A. Since I reactively inhibits A, it confines A's growth, creating a portion of the field with pattern: A in some portion, and I around it, confining it. So even though A and I can diffuse, with diffusion by itself leading to stability of field homogeneity, the combination of reaction and diffusion with faster I and slower A leads to the destabilization of pattern-less homogeneity, and the stabilization of patterned form, morphogenesis.

Turing also knew that reaction-diffusion media can exhibit temporal long-time correlation through *oscillation* of the values of A and I, a system studied in great detail by Hopf (Walgraef, 1997), as well as spatiotemporal correlation through traveling waves of the activator (Cross and Hohenberg, 1993), but he thought that three reactants are necessary to realize time-varying patterns. The more general model we will discuss in this paper establishes how spatial, temporal, and spatiotemporal order could arise in an initially mostly homogeneous medium, without long distance connections or neurons that directly correlate spatially or temporally distant activations (even though such connections can exist and probably influence resulting global patterns).

Meinhardt (1982) and Murray (2001) note that there is one other possible mechanism that can lead to Turing pattern generation: the Substrate-Depletion mechanism. This involves two interacting fields of an Activator (A) and a Substrate (S). A needs S in order to grow, but as A uses up the substrate S, it loses that support depleting S, which results in an inhibition of A. In this reaction, A and S promote A's growth, while A inhibits S. In addition, S degrades itself. Both A and S also diffuse, with S diffusing faster than A for Turing morphogenesis to occur (along with other conditions to be discussed). The result is two differential equations for A and S with reactions $f$ and $g$ which depend on A and S, and diffusion coefficients $D_A$ and $D_I$ as in (1), below, where A and S are fundamental variables, $\partial$ is the partial derivative, and $\Delta$ is the double derivative with respect to space, $\partial_{xx} + \partial_{yy}$, for the diffusion term, which will be applied here to a flat square. The mathematical condition fulfilled for the reaction to be of the Substrate-Deletion variety, as opposed to the Activator-Inhibitor one, is that $\frac{\partial f}{\partial A} > 0, \frac{\partial f}{\partial S} > 0, \frac{\partial g}{\partial A} < 0, \frac{\partial g}{\partial S} < 0$. Since $D_A$ has to be smaller than $D_S$, the latter can be set to unity, while $D_A$can be set to be between 0 and 1.

$$\begin{aligned} \frac{\partial A}{\partial t} &= f(A,S) + D_A \Delta A \\ \frac{\partial S}{\partial t} &= g(A,S) + D_S \Delta S \end{aligned} \quad (1)$$

We propose that the $Ca^{2+}/IP_3$ reaction is an instance of the Substance-Depletion Turing Mechanism. $Ca^{2+}$ is autocatalytic Activator, and is helped by its Substrate $IP_3$, since the latter opens the $Ca^{2+}$ stores, increasing its concentration A. But as $Ca^{2+}$ increases, it effectively depletes $IP_3$, as S, by filling its free enzymatic sites, rendering it less active, effectively inhibiting $Ca^{2+}$. Moreover, as mentioned earlier, just as for the Turing reactants, $Ca^{2+}$ and $IP_3$ both diffuse, but $IP_3$ diffuses faster (Allbritton, et al., 1992).

**Results**: There is an infinite number of systems with specific *f*'s and *g*'s that can realize a Substrate-Depletion system. Our goal is to determine the simplest Substrate-Depletion type Reaction-Diffusion equation that can generate the repertoire of cephalopod chromatic patterns. The search for the simplest model is not due to a belief that natural neuromuscular systems are maximally efficient and non-redundant, since all evidence points to the contrary. As von Neumann (1956) already knew, fault-tolerance and speed can be gained by increase in redundancy of computation. Rather, we seek the simplest system, to know what the *necessary* control degrees of freedom are for generating the static and dynamic patterns must be like, even if actual neuromuscular systems code these degrees of freedom redundantly. If we know this minimum number, we have bounded the possible models from below in complexity. A mathematical approach to model *simplicity* uses polynomial degree as a measure, so the simplest model has the lowest polynomial order, unless the phenomenon to be accounted for requires a higher order. The simplest Taylor-expansion of the functions *f* and *g* in polynomials would yield a linear reaction. However, Ermentrout (1991) and Shoji and Iwasa (2003) have shown that for stripes and spots to be generated in two dimensions, there must be both quadratic and cubic terms. Therefore, knowledge of the repertoire to be generated imposes constraints on the possible differential equations. Taylor expansion of Equation 1, and keeping terms till cubic, was performed by Barrio, Varea, Aragón, and Maini (BVAM) to yield Equation 2 (Barrio et al, 1999):

$$\begin{aligned} \frac{\partial A}{\partial t} &= n(A + aS - CAS - AS^2) + D_A \Delta A \\ \frac{\partial S}{\partial t} &= n(HA + bS + CAS + AS^2) + \Delta S \end{aligned} \tag{2}$$

Where *n, H, a, b, C*, and $D_A$ are coefficients, whose values determine the pattern to be generated. *n* is positive, and $a > 0$, $b < 0$ for a Substance-Depletion mechanism, as will be discussed later. The derivation of the equation starts with Taylor expansion of *f* and g, and curtailing after the cubic term, yielding terms of f and g proportional to: A, S (linear), $A^2$, $S^2$, AS, (quadratic), and $A^3$, $S^3$, $AS^2$, $SA^2$ (cubic). The system is made even more minimal by eliminating the non-interactive quadratic and cubic terms ($A^2$, $I^2$, $A^3$, $I^3$), and leaving only one of two interactive cubic terms (eliminating $IA^2$). As Barrio, et al. (1999) show, there are three fixed points for this system, but if $H = -1$, the system is monostable with a fixed point at A = 0, S = 0. Varea et al. (2007) discuss dynamical patterns possible for $H \neq 1$, like spiral waves, but Leppänen (2004) shows that Hopf bifurcations, and hence dynamic oscillation, are possible even when $H = -1$, so for the rest of this paper, we will only discuss this monostable case, as we believe it is sufficient for the patterns on cephalopod skins. This leaves parameters *n, a, b, C,* $D_A$. Of course, far more complex patterns can be generated as solutions of more general reaction-diffusion equations with Turing and Hopf bifurcations (Richard and Mischler, 2009; Serria-Gonzalez et al., 2021), but our goal is to show that this minimal system is sufficient.

The goal in the coming paragraphs is to determine how the parameters *n, a, b, C,* $D_A$ affect the pattern, and then determining the minimal set, if not the complete set, for generating the entire repertoire. The first step is to determine for which values of the parameters there will be a Turing pattern. Turing (1952) established, using linear stability analysis, the basic conditions in Equation 3 using derivatives of the right-hand side of Equation 1 (Murray, 2001).

$$\frac{\partial f}{\partial A} + \frac{\partial g}{\partial S} < 0$$
$$\frac{\partial f}{\partial A}\frac{\partial g}{\partial S} - \frac{\partial f}{\partial S} + \frac{\partial g}{\partial A} > 0 \qquad (3)$$
$$\frac{\partial f}{\partial A} + D_A \frac{\partial g}{\partial S} > 0$$

Leppänen (2004) shows that the conditions in Equation 3, applied to the BVAM equation, do not depend on *n* and *C*. Furthermore, for each possible value of $D_A$, one could visualize the (b,a) pairs allowing for a Turing pattern as a wedge surrounded by the subsets of the plane in Equation 4. The curves that surround the wedge are bifurcation points.

$$b > \frac{-1}{D_A}$$
$$b < -1$$
$$b < -a \qquad (4)$$
$$a < \frac{(D_A b - 1)^2}{4D_A}$$

Figure 1 shows wedges for $.06 < D_A < .516$. The reason for this range will become clear later. Also, choosing $a > 0$ and $b < 0$ insures that the Turing mechanism is of the Substance-Depletion type. That is because the conditions for this mechanism are that: $\frac{\partial f}{\partial A} > 0, \frac{\partial f}{\partial S} > 0, \frac{\partial g}{\partial A} < 0, \frac{\partial g}{\partial S} < 0$, and for the BVAM equations $\frac{\partial f}{\partial A} = n, \frac{\partial f}{\partial S} = na, \frac{\partial g}{\partial A} = nh, \frac{\partial g}{\partial S} = nb$, so if *a* is positive and *b* negative, the conditions are fulfilled.

The smallest $D_A$ leads to the largest wedge in Figure 1, in bright red, while the largest $D_A$ leads to the smallest wedge in dark blue. At the top of the figure are crosses to show where each wedge begins on the left. Leppänen (2004) also shows that (b,a) pairs below the curve $a < \frac{(D_A b - 1)^2}{4D_A}$ and to the right of $b = -1$ are points at which there is a Hopf pattern of oscillation. We have added an example horizontal line through one of the wedges extending through Turing points and Hopf points to indicate that one could fix the value of *a*, and change the values of *b* only to go from Turing patterns to Hopf patterns, allowing *b* to act as a bifurcation parameter. Therefore, for fixed values of *C, n, and a*, one could switch between stable Turing patterns to oscillating patterns by changing a single parameter b. We believe that this is significant, since pretheoretically, it would seem that static and dynamic patterns are very different types of skin behaviors, and that a central system would need to specify them in drastically different ways. According to the model we have presented, variation of a single parameter *b* is sufficient for switching between static and dynamic patterns, as will be demonstrated later. One issue obvious from Figure 1 is what has been called the *robustness problem* for pattern formation (Woolley et al., 2021): regions of parameter space allowing for pattern are very small. This can be seen in how small the wedges are, especially as $D_A$ rises. We will come back to this issue in the Discussion.

The previous paragraph showed that the parameters *a,b,* and $D_A$ are crucial for determining the kind of pattern that will be obtained, static or dynamic. To understand the contribution of the parameter *n*, we need to consider another aspect of patterns, their spatial frequency *k*. For instance, for a stripe pattern, how many stripes will be packed into the domain? If the spatial frequency *k* is small, there will be a few stripes, whereas if *k* is large there will be many stripes. Turing (1952) thought of the Reaction-Diffusion equation as being similar to a filter, where the frequencies let through are determined by the parameters of the equation. The response of the filter, $\boldsymbol{\lambda}$, has a real part Re($\boldsymbol{\lambda}$), which determines the rate of growth of a spatial frequency, and an imaginary part Im($\boldsymbol{\lambda}$), which determines the oscillatory frequency. Turing (1952) showed that for a Turing pattern to occur, Re($\boldsymbol{\lambda}$) > 0 and Im($\boldsymbol{\lambda}$) = 0. Figure 2 shows Re($\boldsymbol{\lambda}$) and Im($\boldsymbol{\lambda}$) for 3 values of g. For this Figure, a = 1.112 and b = -1.01. The color of the curves indicates the value of $D_A$ starting at $D_A$ = .7 (red) down to $D_A$ = 0 (blue). It can be seen that for very high $D_A$, indicating that A and S diffuse with similar speeds, there is no *k* for which Re($\boldsymbol{\lambda}$) > 0, therefore no pattern develops. When a critical diffusion coefficient $D_A$ is reached at $D_{critical}$ = .528, there is some *k* for which there is pattern, as Re($\boldsymbol{\lambda}$) > 0 and Im($\boldsymbol{\lambda}$) = 0. As $D_A$ continues falling, more and more spatial frequencies are allowed, raising the bandwidth of the pattern, till a level is reached at which the pattern is *highpass,* in filter terms. Note also that for the chosen values of *a* and *b*, Im($\boldsymbol{\lambda}$) = 0 at the values of *k* leading to pattern. Each column in Figure 2 is for a different value of *n*. Note that as *n* increases, the critical spatial frequency $k_c$ at which pattern is possible is higher. We therefore arrive at the physical meaning of *n*, which conditions the spatial frequency of the pattern.

The last parameter is *C*, the coefficient of the quadratic term. Since the coefficient of the cubic is 1, C is understood as a weight on the significance of the 2 nonlinear terms, quadratic vs. cubic (Ermentrout, 1991; Shoji and Iwasa, 2003). There are two methods for understanding the contribution of the nonlinear terms: weakly nonlinear analysis and numerical methods (e.g., Leppanen, 2004). We have opted for the numerical route, since we found weakly nonlinear analysis to face problems as the perturbation size increases (Leppanen, 2004), and to be mathematically quite complex, requiring many assumptions. For numerical simulation, we used the method of lines and the Tensor Product Grid Method (Fornberg, 1998), utilizing the **NDSolve** engine in *Mathematica*® on 101x101 grids (details in Appendix A). Since our goal is determining the fewest number of parameters for generating the repertoire, we fixed as many parameters as possible. The parameters *a*, *b* were fixed at *a* = 1.112 and *b* = -1.01 to generate Turing patterns, and *n* = 0.4 for an intermediate critical spatial frequency. The values for *a* and *b* were chosen on a horizontal line such as the one in Figure 1, where increasing *b* can lead to a Hopf pattern, as will be discussed later. It's important to note that other values for these parameters could have been chosen, as long as the (b,a) pair is in one of the Turing wedges in Figure 1. Even though we manipulate the coefficient C of the quadratic term, it is also possible to manipulate the cubic terms as in Woolley et al. (2021).

Besides varying *C*, we also varied $D_A$. Ermentrout (1991) and Shoji and Iwasa (2003) showed that varying the prominence of the quadratic vs. the cubic terms, here via *C*, would lead to stripes when the cubic dominates (*C* near 0), and spots would be produced when the coefficients are more equal (*C* near 1). We also predicted that $D_A$ would be important for the fineness/stochasticity of the pattern and that very low $D_A$ would lead to less disruptive patterns, whereas the higher $D_A$, the more disruptive the pattern. The reason is that low disruptive mottle-type patterns show correlation on small patches only, requiring the extent of diffusion to be small, which is what happens for low $D_A$. And the opposite is the case for disruptive patterns with large structures like stripes and spots. The first columns of Figures 3 and 4 provide example skin patterns from different cephalopods, while the second columns show simulations. Values used for the simulations are in Table 1. The initial condition of the field is gaussian random noise to displace the initial condition slightly away from the fixed point. Each time a simulation is run, therefore, the exact pattern will be slightly different, but will be within the class represented. The match of the simulation and natural pattern are not intended as exact but as suggestive of the class. And we feel that a quantitative fit of the example patterns and simulations is not of import, since there can be enormous variation in the real patterns.

Figure 5 provides a sample of patterns obtained by varying $C$ and $D_A$: $0 < C < 1$ and $.06 < D_A < .48$. Based on our previous observation and description of cephalopod body patterns (Mather and Alupay, 2016), we believe that any of the simulations we observe in Figures 3-5 could be observed on the whole skin of a cephalopod or a patch thereof. And as expected, as $C$ increases, we transition from stripes to spots, and as $D_A$ increases, there is a transition from less disruptive/non-conspicuous patterns to ones that are more disruptive/conspicuous, mottle through blotch/reticulate, to stripes and spots. We believe that the qualitative correspondences in Figures 3-5 add hope for the prospect of a mathematical theory underlying the ethology of cephalopod skin displays. Such theory would allow comparison amongst very different animal behaviors, such as speech production, to which we have applied this very same theory (Iskarous, 2019).

The third column of Figures 3 and 4 provide the log-polar transforms of 2D Fourier spectrum (2DFT) of each pattern, which is described in Appendix B. The 1D spectrum has been a very useful quantitative device in the empirical research on cephalopod patterns (Chiao, et al., 2010), and we believe that the additional information in the 2DFT (Zylinski et al., 2011) furthers the use of spectra in the description and comparison of patterns. The log-polar transform is used, since it neutralizes the direction of the patterns. For instance, the 2DFT of stripes in different directions will consist of 3 dots on the unit circle, with the lowest-angled one specifying the angle of the stripes. Log-polar transformation effectively integrates out the direction of the stripes. Low spatial frequency is on the top left and frequency increases outward. As can be seen from the third column of Figures 3 and 4, lines and spots have peaks at relatively low frequencies, and exhibit very little high frequency energy. Irregular patterns have higher frequencies excited. This agrees with the results of Hanlon et al. (2009) in that it is possible to spectrally distinguish the more *disruptive* patterns like lines and spots as low frequency from the less disruptive ones like mottle, which are higher frequency. The spectral understanding is furthered in the 2DFT, however, since we can distinguish nicely *across* and *within* the classes. And as explained in Appendix B, the 2DFT clarifies the superpositional structure of displays, investigated by Anderson et al. (2003) and Kelman et al. (2006), allowing us to see spots as a superposition of stripes, for instance, which goes a long way to explaining why the number of degrees of freedom underlying the patterns is very small.

The large difference between the patterns of the static repertoire makes it surprising that just two parameters are necessary for generating them. To provide support for the low degree freedom nature of the patterns, we simulated 60x60 patterns, again varying in $C$ and $D_A$, and we applied the Laplacian Eigenmaps nonlinear manifold reduction method for finding low-dimensional representations (Belkin and Niyogi, 2003). We first performed the log-polar transform 2DFT on patterns, so that stripes of different direction, for instance, would become more similar in the transform. We presented the 3600 patterns, each as as 101x101 = 10201dimensional vector of Log-polar 2DFT coefficients, and specified a maximum number of 5 dimensions for the reduced representation. Figure 6 shows the first coordinate as a function of the $C$ and $D_A$ of the pattern. By viewing the 2D simulated patterns responsible for specific $C$ and $D_A$ and by using Figures 3-5, we labeled regions of Figure 6 with the ethological names for patterns that would be traditionally used. This further proves the low-dimensionality of the skin patterning process, since both synthesis and analysis lead a low-dimension. This supports the work of Ishida's (2021) using Cellular Automata and Convolutional also on static patterns of cephalopods, work that has been concurrent with ours.

Therefore, for static patterns, while $a$, $b$, and $D_A$, specify the type of pattern, Turing vs. Hopf, and $n$ specifies the overall density of structures, $C$ and $D_A$ jointly determine the disruptiveness and spot/stripe nature of the pattern. To simulate dynamical patterns, such as oscillation, with changing as few parameters possible, we simply increased $b$. Figure 7a shows the Re($\boldsymbol{\lambda}$) and Im($\boldsymbol{\lambda}$) as $b$ is increased, when $k = 0$. The reason for concentrating on $k = 0$ is that in oscillating across the whole medium, the lowest spatial frequency is involved. Hopf oscillation in such a medium occurs when Re($\boldsymbol{\lambda}$) $> 0$ and Im($\boldsymbol{\lambda}$) $> 0$ at k = 0. From Figure 7a, it can be seen that this happens at b = -1. Therefore, to simulate an oscillatory pattern, we kept $a = 1.112$, $g = .4$, $D_A = .45$, $C = 1$, and $H = -1$, and only increased $b$ to -0.93. Figure 7b shows the mean pixel intensity as a function of time for 2 simulations, one of a Turing pattern with $b = -1.01$ (green) and the other of a Hopf pattern $b = -0.93$ (magenta). The Turing mean intensity gets to a stable value and stays there, while the Hopf mean intensity oscillates. We also provide a movie of this simulation (Video 2, Supplementary Information). This oscillation is similar to what can be seen in How et al. (2017, Video 3.1.1, Supplementary Information). As can be seen in the movie and Figure 7b, oscillation of a relatively fixed magnitude is achieved at about 30% into the simulation period. Examination of the transient first portion of the movies shows asynchronous oscillation, which How et al. (2017) call flickering. Figure 7c shows the intensity of three randomly chosen pixels in the simulation. It can be seen that at the beginning, 2 of them (black and green) are in phase, while one (red) is out of phase with the first two. As the simulations proceed, all of them proceed to being in phase. Therefore, we believe that asynchronous and synchronous oscillation could be realized at the same parameter values, but that asynchronous oscillation is a transient on the way to synchronous oscillation. Moreover, of we keep $b = -.93$, and change the initial conditions from random gaussian noise to a strip of value 1, the strip will translate, reminiscent of the passing cloud (Mather and Mather, 2004) traveling waves. The lowest row of Figure 7 shows a few frames from such a simulation, where a strip translates forward and backward, which is like the multidirectional waves observed by How et al. (2017). With the static patterns, we saw that each has a different parameter setting, but many dynamic patterns can be achieved with the same setting, but with a change in initial conditions or the stage of the action.

**Bimodal Population Codes:** Deterministic differential equations describe a medium where each location is updated synchronously with all other locations, and noise plays a role only in the initial conditions. We know that this is not the way that neurons in a brain and muscle cells in a syncytium would operate, as they are fundamentally probabilistic, with each unit having a probability to fire or not, and different units fire asynchronously (Cowan, 1967). For the last 30 years, Cowan and his colleagues have developed stochastic dynamical frameworks based on the mathematics of Quantum Field Theory to model asynchronous noisy neural fields (Cowan, 2014). We present here a much simpler approach for associating a nondeterministic asynchronous model to a Turing deterministic synchronous model. This method is based on a method developed by Marquez-Lago and Padilla (2014).

To appreciate this method, we start with establishing the use of bimodal probability distributions to describe skin displays. If we imagine the intensity of A on the x-axis of a graph, e.g., as in Figure 8a or 8b, with the y-axis representing the probability of that intensity A, a bimodal distribution would indicate that there are two intensities that are quite stable and highly probable. In the blue curve of Figure 8a, one intensity of A = 1, which is dark/activated, is highly probable, while another intensity, -1 light/unactivated, is also probable, but less so. If this blue probability curve is interpreted as the probability distribution for the intensity of activation of *any* chromatophore's muscles, then the most probable chromatophore intensity would be A = 1, so most chromatophores would be dark, and therefore we would say that the *background* brightness is dark. The less probable A value would be the less frequent, which if we add the additional constraint that nearby chromatophores have similar activation, would therefore be the brightness of *foreground* light spots on the dark background. Note that more probable is identified as background, and less probable as foreground (cf. figure/ground distinction in perception). Therefore, if the instruction to muscle cells is to fire or not independently based on the blue probability distribution in Figure 8c, and we add the additional constraint of common activation of nearby chromatophores (through diffusion), then the observed global pattern would be dark background with light spots (Figure 3d). If the two probability modes become more and more equal, as in the red curve in Figure 8a, the distinction between foreground (less probable) and background (more probable) is blurred, and we get stripes.

Marquez-Lago and Padilla (2014) draw a relation between three domains: 1) quadratic vs. cubic term balance in a reaction-diffusion equation, which uses the parameter *C*; 2) Double-well potential wells, which are the negative integrals of quadratic and cubic terms, and are thereby quartic polynomials; 3) Bimodal probability distributions obtained by taking the negative exponential of the double-well. Under this association, varying the parameter *C* changing the balance of quadratic to cubic terms in a reaction-diffusion differential equation corresponds to changing the depth of one well with respect to another of a double-well potential, which then corresponds to changing the heights of modes in a bimodal distribution, a Boltzmann Distribution. And such a distribution can be associated with skin displays as we outlined in the previous paragraph.

Marquez-Lago and Padilla (2014) show a way of translating from reaction-diffusion systems to Fokker-Planck (FP) equations whose solutions are Boltzmann distributions. The prediction is that if *C* is near 0 in BVAM, the associated FP solution would have 2 modes of roughly equivalent heights, and as *C* moves towards 1, one mode would become much higher than the other. We tested this hypothesis by solving FP with the values for the patterns simulated in Figures 3 and 4, found in Table 1. The Langevin method of simulating FP was used on the BVAM Equations (2), via Euler-Maruyama discretization (details in Appendix C). In these simulations, we solve a Langevin stochastic differential equation with noise determined by the diffusion coefficients and bias determined by the reaction terms (Rajotte, 2014), and the probability distribution of A after 500,000 times steps (dt = .01) is recorded for 5000 simulations of the same stochastic differential equation. The resulting histograms of A at convergence as *C* and $D_A$ are varied in Table 1 are in the fourth columns of Figure 3 and 4. As expected, the spot-like patterns have large skewness, while the line-like patterns have lower skewness. Figure 8c shows the calculated Pearson Skewness histograms from Langevin-simulations with fixed $D_A$ and varying *C* from -1 to 1. The skewness of the empirical histograms of pixel intensity in the reaction-diffusion simulations such as those in Column 2 of Figures 3 and 4 is also shown. Both vary from high to low skewness, as expected, but one surprising result was that the Langevin-based skewness crossing point was at C = -.4, not C = 0, as expected. We have eliminated the possibility that this is due to the specific $D_A$ values used. We believe that this discrepancy is due to the linear components of BVAM, but we leave this for future investigation.

We interpret the parametrizable bimodal distributions to be population codes (Dayan and Abbott, 2001). In such a code, each motor state of the system has a probability of being realized. Here the motor states are the degree of activation of muscles of a single chromatophore or a small population thereof. Neurons fire or not according to these bimodal distribution with centrally-specified parameters like skewness C that specify the relative heights of the modes. There is an additional constraint that small populations fire similarly to their neighbors, which is mathematically similar to the diffusion effect. Most of the literature on population codes focuses on single-mode distributions, but bimodal distributions have also been investigated theoretically (Zemel et al., 1998). And Amirikian and Georgopulos (2000) provide empirical evidence of bimodal population codes for motor cortical cells in monkeys in reaching behavior.

**Discussion and Conclusion**: In this paper we have provided a theory for framing the Horizontal Control Hypothesis, based on recent neurophysiological and structural data suggesting that the Hypothesis is a possibility (Rosen, 2016; Senft et al., 2021). We have sought to provide the simplest concrete instantiation of those rules via Turing and Hopf pattern formation theory using the BVAM universal simplification of nonlinear reaction-diffusion equations. We have shown that the Horizontal Control Hypothesis specifies interactional parameters for each location in the medium. We have given meaning to these parameters, and showed that the minimal set of alternating coefficients is 3: *b* for specifying static vs. dynamic, and *C* and $D_A$ for specifying the spot/stripe aspect and the disruptivness. *a* and *H* could be fixed. It will be interesting to see if inclusion of more effects such as the curvature of the domain will require this low number of varying parameters to increase.

We have also provided a biochemical interpretation of the reaction-diffusion equation in terms of $Ca^{2+}/IP_3$ dynamics. This generates testable predictions, since the coefficients have physical meaning. For instance, the parameter *b* is the rate of degradation of $IP_3$, which can happen in many ways. Is this rate actively controlled by a cephalopod if it's trying to change from a fixed to an oscillator pattern? *a* and *H* stand for the linear aspect of the cross-effects of $Ca^{2+}$ and $IP_3$ on each other. Are those effects relatively fixed for an animal? One of the roles of computational neuroscience is to provide theories for generating space of predictions, such as these, but it is of course up to empirical research to answer the questions. The Marquez-Lago and Padilla (2014) interpretation of the *C* parameter further enhances our neural interpretation of the horizontal control hypothesis for the static patterns by suggesting that what is centrally specified is a bimodal probability distribution for firing, together with a constraint that nearby units fire together. Senft et al. (2021)'s recent evidence of axo-axonic synapses could have import on the probablistic interpretation we provide, since it's well-known that these synapses do not lead to action potentials, but the modulation of the probability of firing (Howard et al., 2005). We also believe that the present work has import for the empirical study of cephalopod patterns (Osorio, 2014) as the higher-level interaction parameters can be used to robustly quantify pattern similarity and differences. We also believe that the work has implications for electrophysiological and biochemical work on cephalopod chromatic displays, as the reaction-diffusion mechanisms we have discussed could guide the search for the biochemical events instantiating them, as has happened in the study of developmental fish skin patterning (Asai et al., 1999).

One issue that we have mentioned earlier is the robustness problem (Woolley et al., 2021), regarding the small regions of parameter space where pattern can occur. This is true mathematically, however we believe that lack of robustness has an advantage for the animal: flexibility. If the region in parameter space realizing a pattern is very large, the physical implementation would require enormous dynamic range for the involved biochemicals. Such range is important for dynamical states, such as the concentration of $Ca^{2+}$, for instance, but far less so for the architectural parameters of a system. Keeping the range of such parameters relatively small allows for easy switching between different patterns, by having small regions in bifurcation space, which is exactly what we see on the cephalopod skin, which requires fast flexible switching.

In earlier work, Iskarous (2019) used the BVAM equations to derive the macroscopic patterns in another behavior in an animal widely separated in evolution from cephalopods: speech in humans. In the latter, the patterns are the gestures of the tongue that specify the consonants and vowels of a language. That work shows that a neural system operating under a BVAM system of equations with Turing and Hopf bifurcations can generate several empirically verified signatures of vocal tract action. Our work on cephalopods is inspired by dynamical work on speech production in the Articulatory Phonology framework (Fowler et al., 1980; Browman and Goldstein, 1989; Saltzman and Munhall, 1989). This is part of even more general work on how animals control the abundance of degrees of freedom they have at their disposal (Bernstein, 1967; Turvey, 1990; Latash, 2012). This work on cephalopods provides a nice example, where the patterning to be explained is extremely clear, and we feel that it will allow further progress in our understanding of speech gestures, for example, despite the large surface differences in behaviors. A common mathematical vocabulary for different animal behaviors helps the cross-fertilization of methods and results across widely different fields of animal behavior, such as the locomotion of nematodes and gaits of many animals (Golubitsky et al., 2004). Advances in mathematical biology, which started with Turing, have had major influence on our understanding of animal development. We believe that his work could also guide the development of a of minimal mathematical language for describing the vast diversity of animal action, perception, and perhaps even cognition, since in all of these there are local computations and global patterns.

**Conflict of Interest**: The authors have no conflict of interest to declare.

**Acknowledgements**: This research was possible due to funding by NSF grant 1246750. We would like to thank Bill Kier, Tatiana Leite, Louis Goldstein, Alessandro Vietti, and the late Roland Anderson for helpful discussions.

Captions

**Figure 1**: Wedges of Turing patterns and Hopf patterns for different values of $D_A$, which varies from small (red) to large (blue) through a rainbow color encoding.

**Figure 2**: Rate of growth, Re($\boldsymbol{\lambda}$), and frequency of oscillation, Im($\boldsymbol{\lambda}$), as a function of spatial frequency $k$, for 3 values of $n$.

**Figure 3**: Disruptive Cephalopod skin displays. Column 1: Pictures of displays on the skins of cephalopods. Column 2: BVAM reaction-diffusion simulations of the displays. Column 3: Log-polar transforms of Two-dimensional Fourier Spectra of the simulations. Column 4: Simulations of Fokker-Planck stochastic versions of the BVAM equations for studying pattern selection. Row 1: Lines (*Octopus chierchiae*; Photo: Roy Caldwell). Row 2: Irregular Stripes (*Octopus chierchiae*; Photo: Roy Caldwell). Row 3: Stripes and spots (*Octopus chierchiae*; Photo: Roy Caldwell). Row 4: Light on Dark Spots (*Sepia pharaonis*; Photo: Authors). Row 5: Dark on Light Spots (*Sepia pharaonis*; Photo: Authors) (Color Online).

**Figure 4**: Nondisruptive Cephalopod skin displays. Column 1: Pictures of displays on the skins of cephalopods. Column 2: BVAM reaction-diffusion simulations of the displays. Column 3: Two-dimensional Fourier Spectra of the simulations. Column 4: Simulations of Fokker-Planck stochastic versions of the BVAM equations for studying pattern selection. Row 1: Reticulate (*Octopus chierchiae*; Photo: Roy Caldwell). Row 2: Light on dark Blotch (*Abdopus sp.*; Photo: Authors). Row 3: Dark on Light Blotch (*Octopus insularis*; Photo: Tatiana Leite). Row 4: Dark Mottle (*Octopus chierchiae*; Photo: Roy Caldwell). Row 5: Light Mottle (*Abdopus sp.*; Photo: Authors). Row 6: Uniform Light (*Abdopus sp.*; Photo: Authors). Row 7: Uniform Dark (*Octopus chierchiae*; Photo: Roy Caldwell) (Color Online).

**Figure 5**: A sample of patterns varying in $C$ (rows) and $D_A$ (columns) (Color Online).

**Figure 6**: Values on dimension 1 of a Laplacian Eigenmap dimensionality reduction over log-polar 2DFT's of 60x60 patterns varying in $C$ and $D_A$ (Color Online).

**Figure 7**: Rate of growth, Re($\boldsymbol{\lambda}$), and frequency of oscillation, Im($\boldsymbol{\lambda}$), at $k = 0$, as a function of $b$ (panel a). Oscillation of the mean of A across frames in a Hopf pattern (panel b).

**Figure 8**: Bimodal Distributions, with highest peak corresponding to background and lower peak to foreground patches, where the background can be dark (a) or light (b). The skewness of bimodal Boltzmann distributions resulting from Langevin-simulations of the Fokker-Planck version of BVAM (red) and the deterministic simulations (blue) (c) (Color Online).

**Table 1**: Values for BVAM parameters for the static and dynamic simulations that generate the different types of pattern

Figure 1

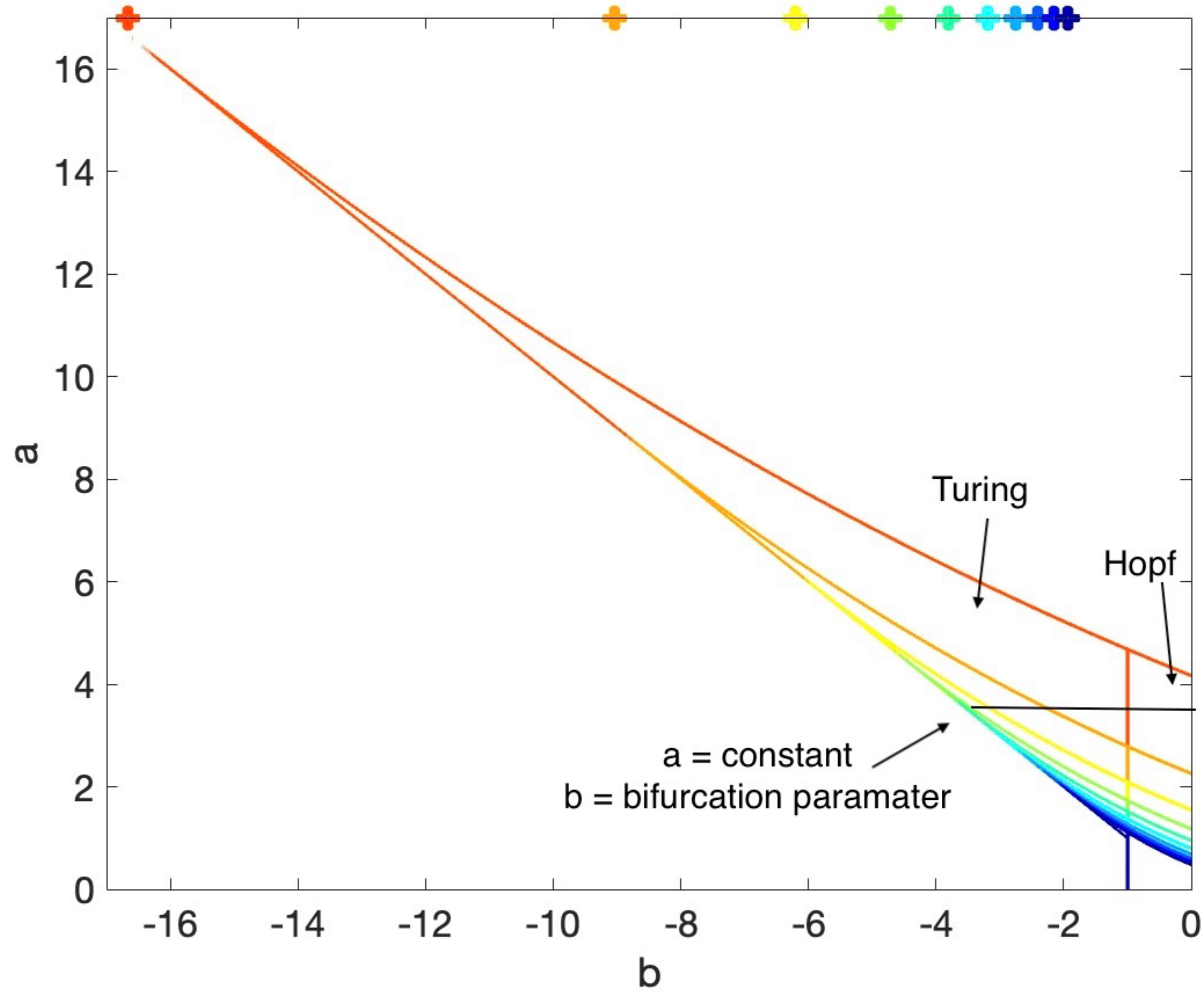

Figure 2

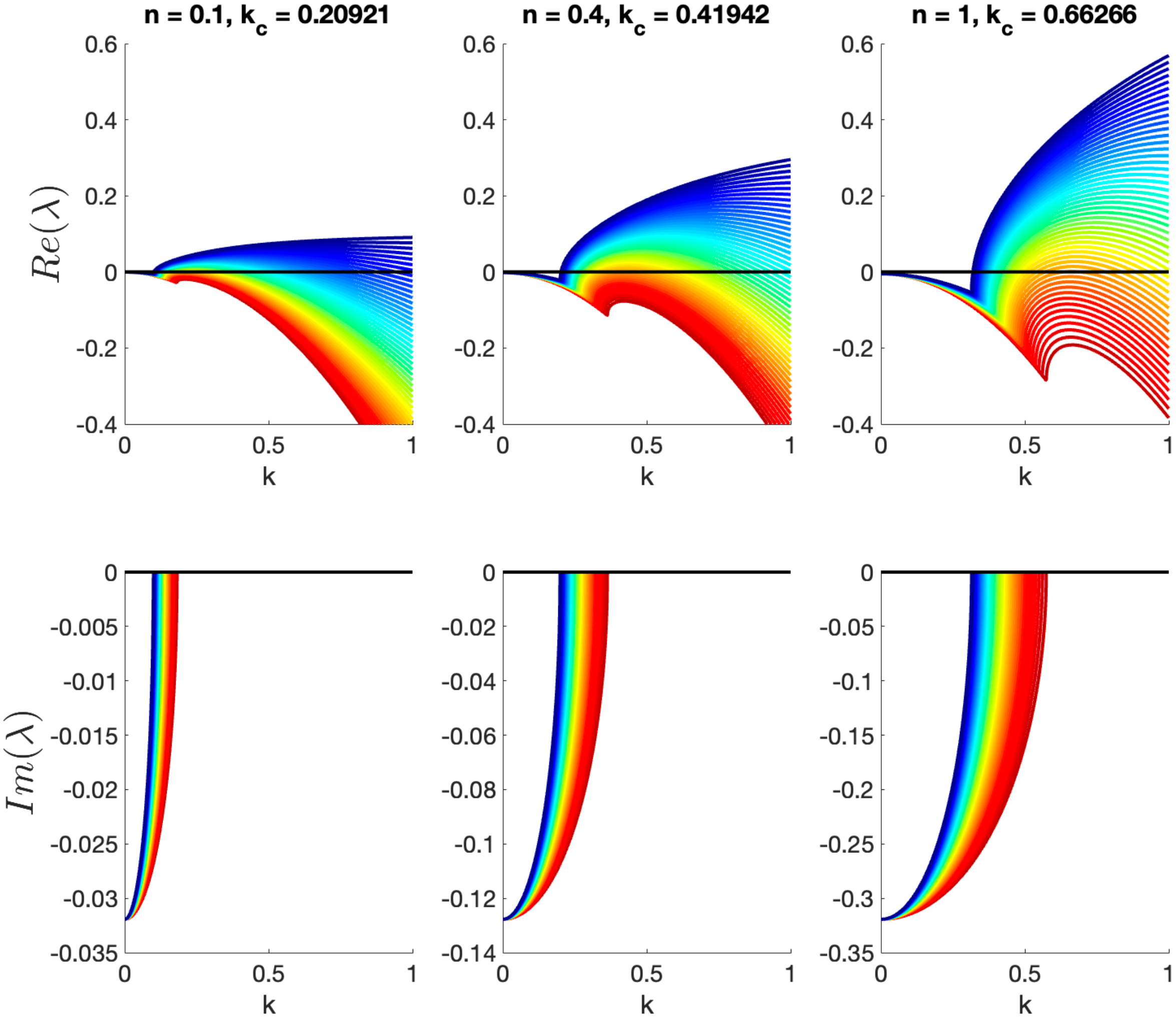

Figure 3

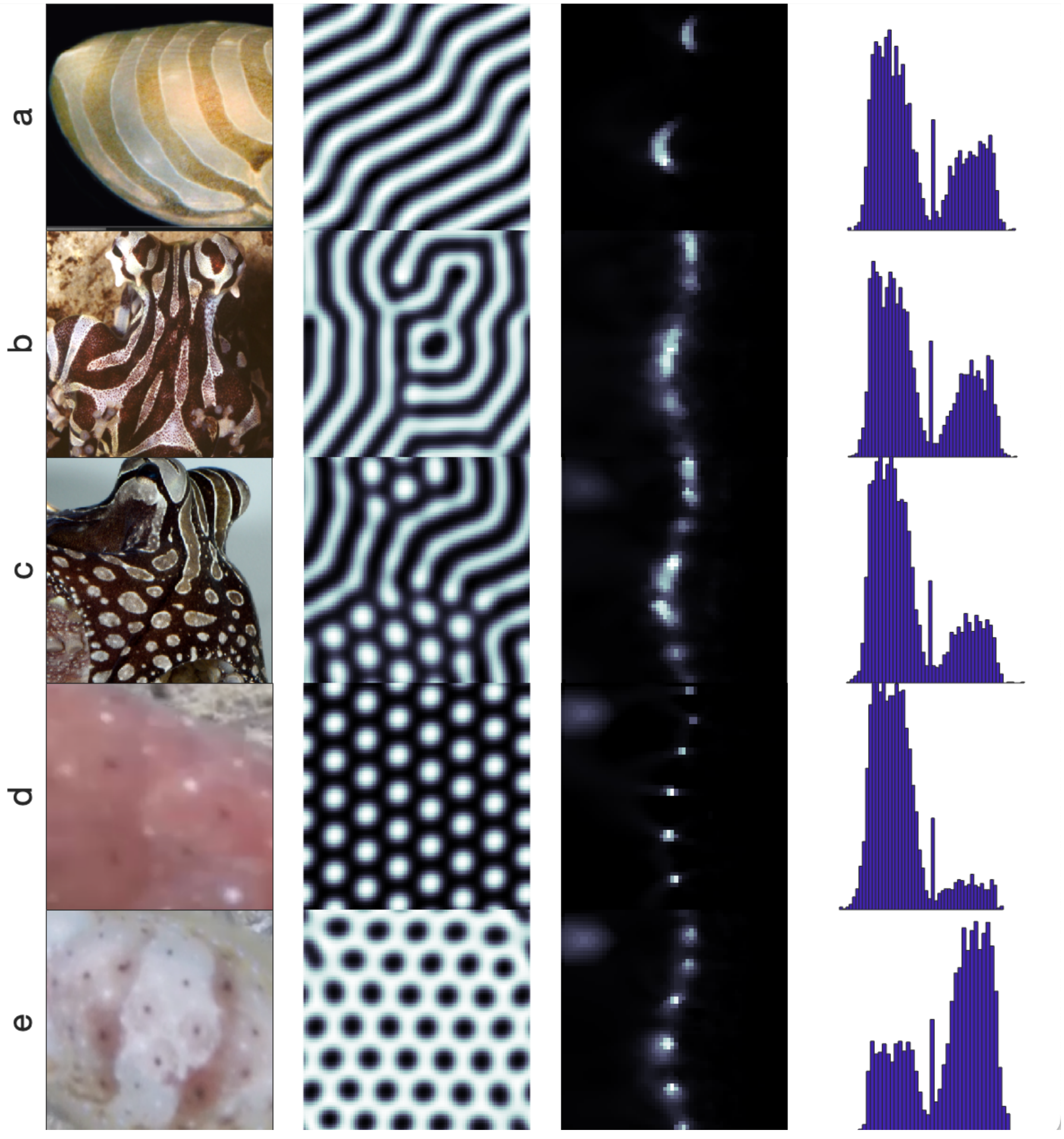

Figure 4

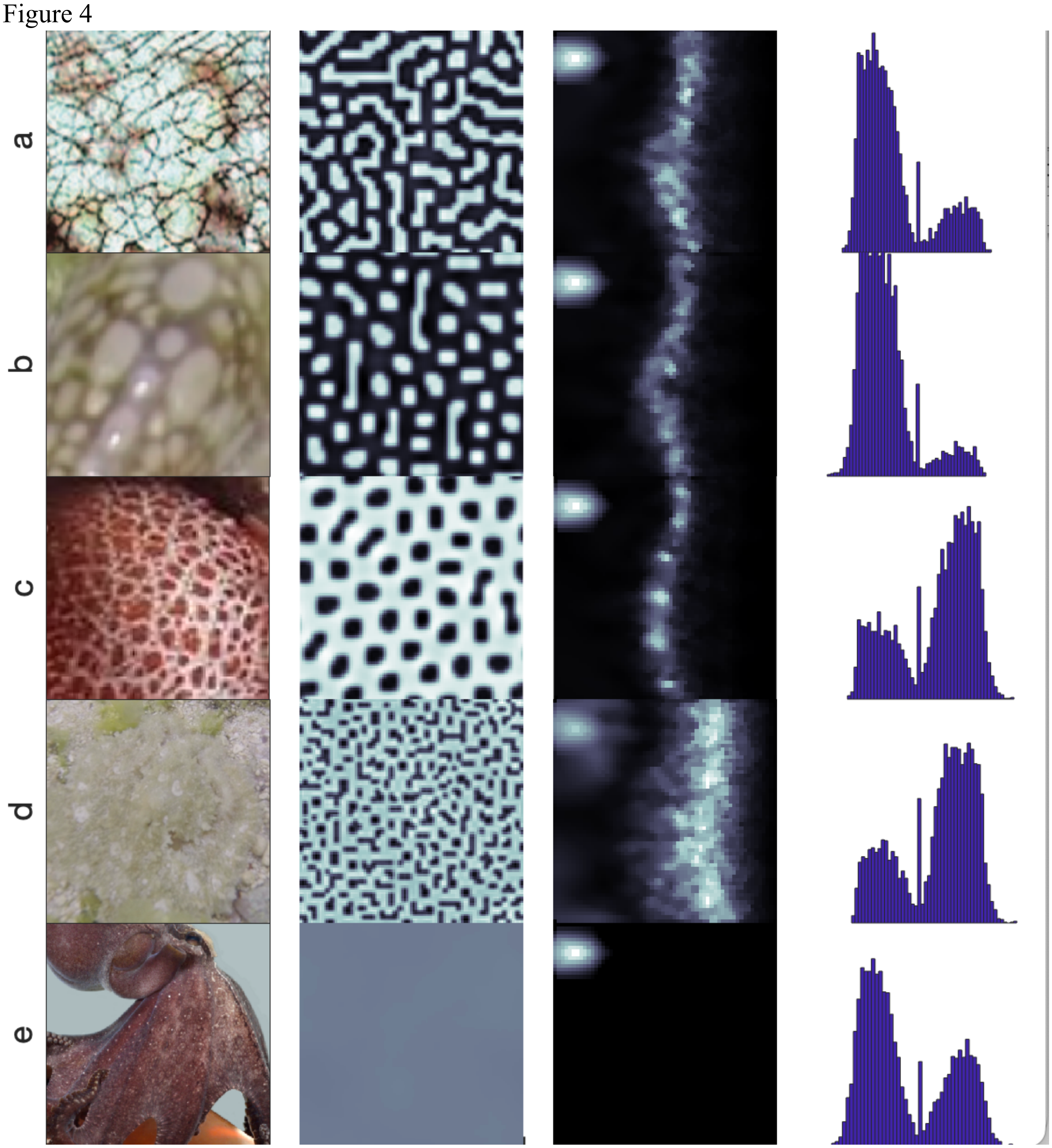

Figure 5

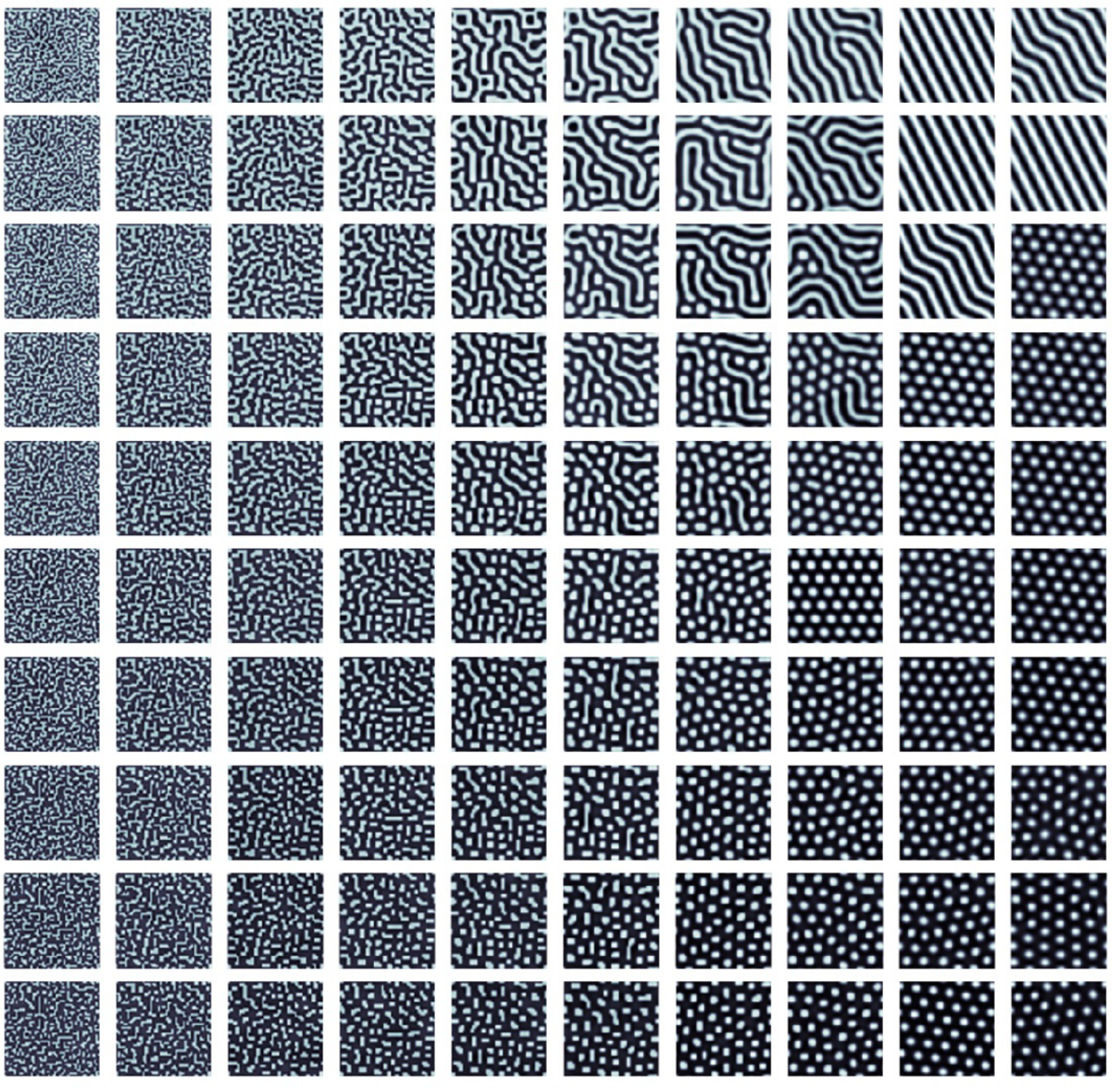

Figure 6

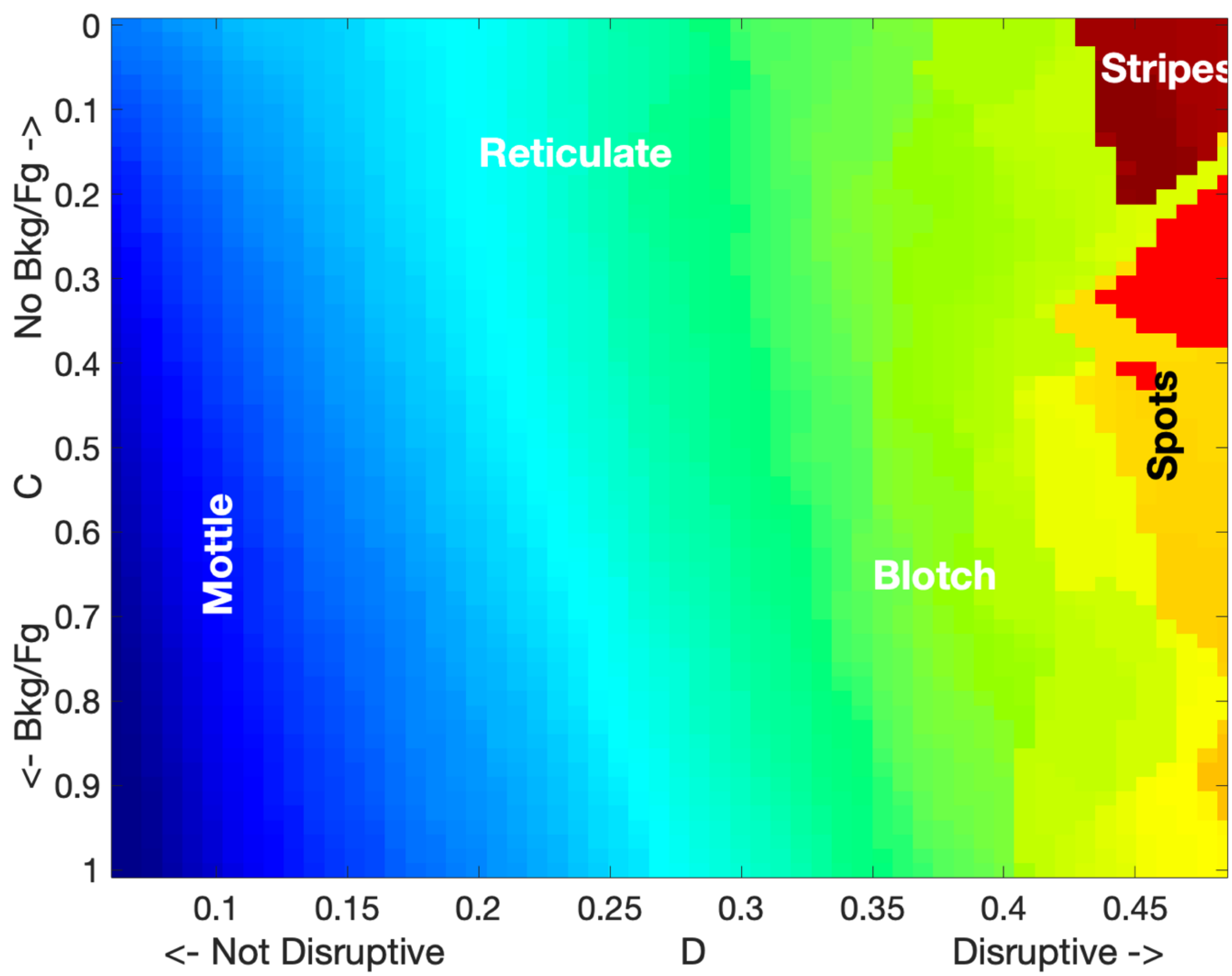

Figure 7

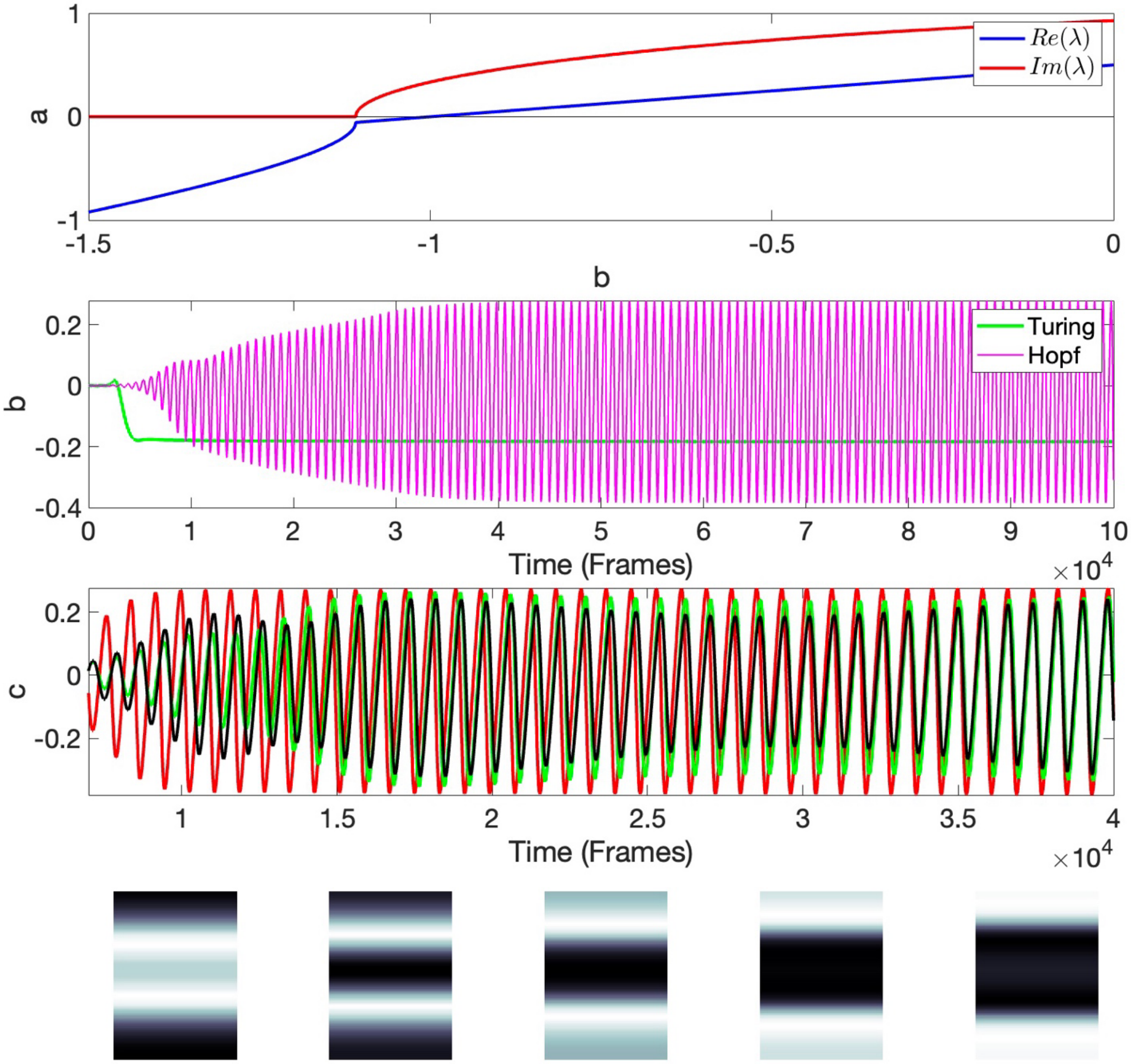

Figure 7

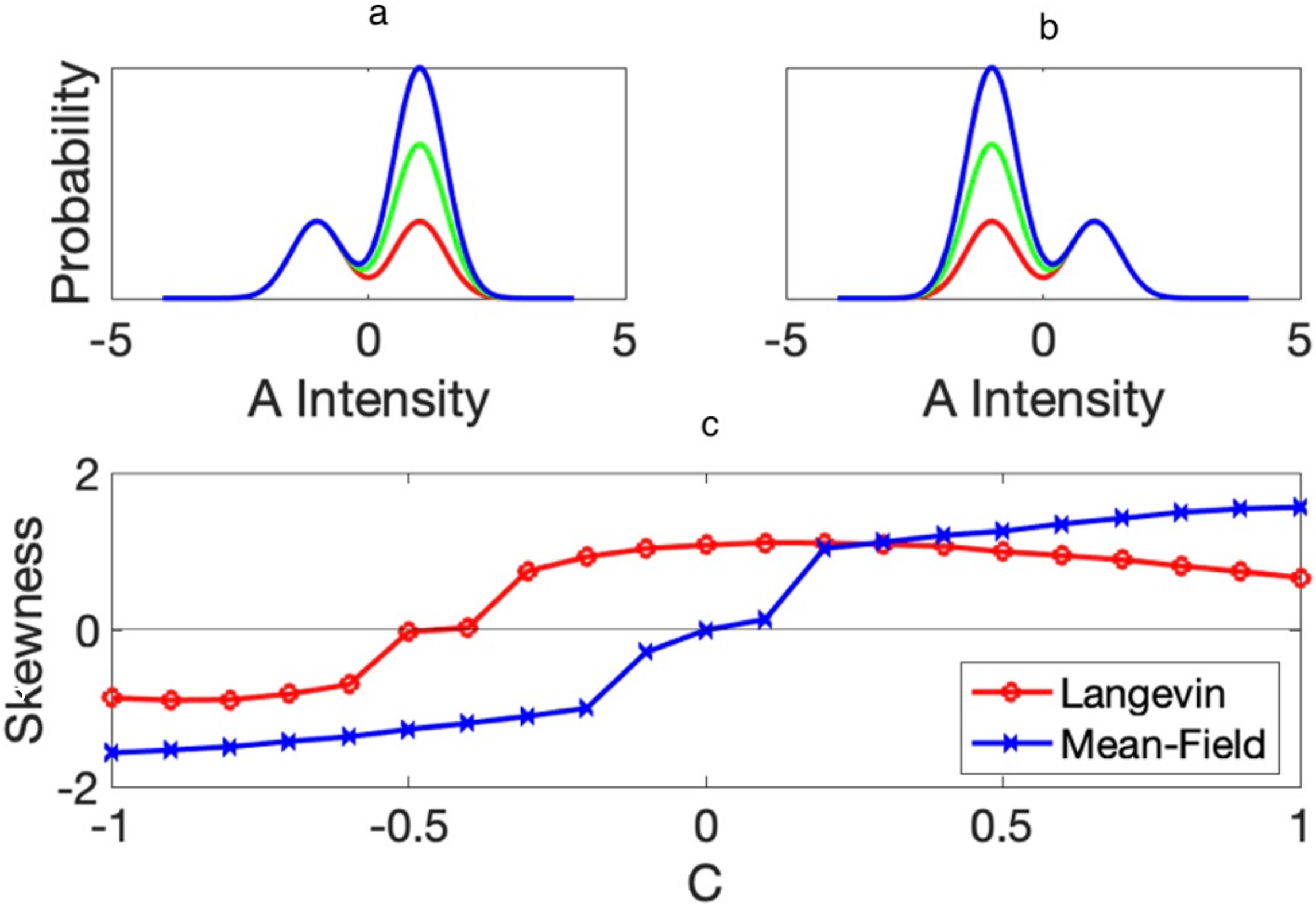

Table 1

| | Figure | a | b | g | C | H | $D_A$ |
|---|---|---|---|---|---|---|---|
| Stripes | 3a | 1.112 | -1.01 | .4 | 0 | -1 | .455 |
| Irreg. Stripes | 3b | 1.112 | -1.01 | .4 | 0 | -1 | .4 |
| Stripes + Spots | 3c | 1.112 | -1.01 | .4 | .25 | -1 | .45 |
| Light Spots | 3d | 1.112 | -1.01 | .4 | .5 | -1 | .45 |
| Dark Spots | 3e | 1.112 | -1.01 | .4 | -.5 | -1 | .45 |
| Reticulate | 4a | 1.112 | -1.01 | .4 | .3 | -1 | .2 |
| Light Irreg. Spots (Blotch) | 4b | 1.112 | -1.01 | .4 | .8 | -1 | .35 |
| Dark Irreg. Spots (Blotch) | 4c | 1.112 | -1.01 | .4 | -.8 | -1 | .35 |
| Mottle | 4d | 1.112 | -1.01 | .4 | -.8 | -1 | .06 |
| Uniform Dark | 4e | 0 | 0 | 0 | 0 | 0 | 1 |
| Oscillation | 7 | 1.112 | -.93 | .4 | 1 | -1 | .45 |

## Appendix A: Numerical Reaction-Diffusion Simulations

Numerical simulations were performed using Mathematica®'s NDSOLVE engine, which uses the method of lines. Spatial derivatives are approximated using a Tensor Product Grid Method (Fornberg, 1998). The step size for the time derivative is chosen adaptively. We used a 101x101 grid size, and the maximum time was 200,000. We used periodic boundary conditions, which is unusual, since most previous studies have used zero-flux boundary conditions. de Witt (1996) showed, however, that unless the dx is very small, which is problematic for stability, zeros-flux boundary conditions introduce significant artifacts to the pattern. Moreover, in viewing cuttlefish dynamic patterns, it often seems that waves that pass through one edge of the animal, reappear on the opposite edge, so periodic boundary conditions may be more appropriate. For the simulation movie for the dynamic pattern, we could not use the method above, since the time step is adaptive, therefore for that, we used Euler's backward method with dt = .05.

## Appendix B: 2D Fourier Transform

The one-dimensional Fourier transform centers around sinusoids, and tries to find the frequency context of signals. The top left panel of Figure SIB1 shows a sinusoid with three peaks (3Hz) on a spatial domain x. If a 2D spatial domain has that same sinusoidal variation on coordinate x, but is constant, for each x, along y, then we can represent the variation in the amplitude using the top right panel. Seen as a relief, we get the bottom left display, where dark color is high amplitude, and light color is low amplitude. In one dimension, the Fourier spectrum will have one positive and one negative peak. The 2D Fourier Spectrum, also called k-space, shown in the bottom right panel, will have peaks in the x-direction, seen as two light spots. But it will not have peaks in the y-direction, as there is no variation in y. Note that stripes, or lines, are the simplest kind of pattern from the 2D Fourier perspective, in the same way that a sinusoid is the basic signal in 1D.

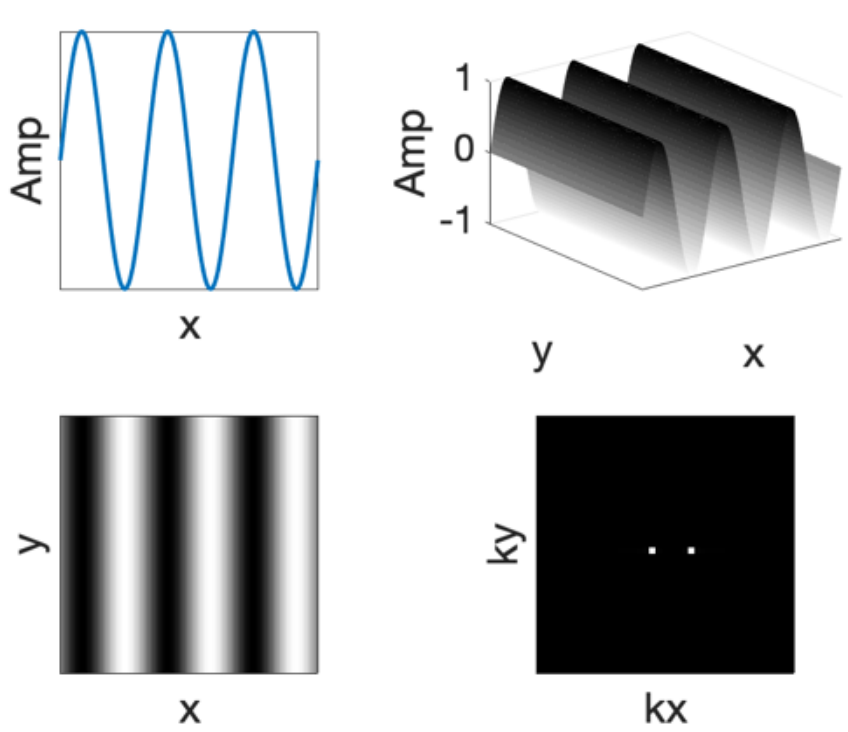

Figure SIB2 top row shows the pattern (left) and 2D Fourier Transform (right) for stripes again. If the spatial frequency of stripes rises, as seen in row 2, the 2D spectral peaks travel outwards. If the stripes are tilted, so are the peaks in the spectrum, as seen in the third row. Therefore, the radial location in the 2D spectrum represents the frequency of variation, whereas the angular direction represents the tilt of the pattern. More complex patterns can be built from stripes. So, if three sets of stripes at 0 degrees, 60 degrees, and 120 degrees are added, the superposition of

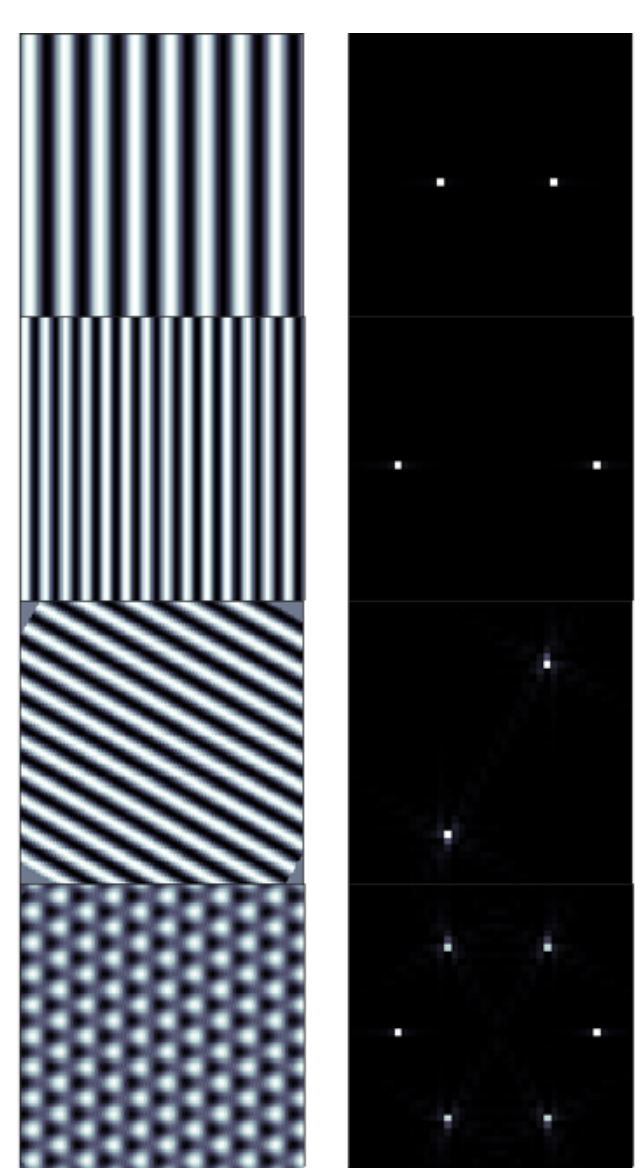

the peaks and troughs would align in such a way as to produce spots, as in the lowest row. It can be seen from this exposition that as D falls in the equations, the 2D spectrum will have more and more peaks as we travel outwards. As C rises in magnitude towards -1, the more peaks there will be near some circle. Therefore, the Fourier 2D spectrum, which can be easily computed from real patterns, directly reveals the two main parameters, C and D, of the underlying reaction-diffusion equations. As discussed in the text, the log-polar transform of the 2DFT allows for two stripe patterns in different directions to have equal transforms.

## Appendix C: Fokker-Planck Simulation

To obtain the pattern-selection histograms in the fourth column of Figures 1 and 2, the Fokker-Planck equations were simulated via the Langevin method. In this method, the BVAM equations become:

$$dA = g(A + aI - CAI - AI^2) + D_A \zeta_A \sqrt{dt}$$
$$dI = g(HA + bI + CAI + AI^2) + \zeta_I \sqrt{dt}$$

Where $\zeta$ is white noise. So, space has disappeared, and has been replaced by a gaussian noise process whose variance is the diffusion coefficient, and whose bias is the reaction portion of BVAM, with dt = .01. For each panel of column 4 of Figures 1 and 2, 10000 simulations were done, and the panels show the histograms of A for each combination of parameters.